# Role of $H_2O$ and $O_2$ Molecules and Phosphorus Vacancies in the Structure Instability of Phosphorene

Andrey A Kistanov[1,2], *Yongqing Cai[2], †Kun Zhou[1], Sergey V Dmitriev[3,4] and ‡Yong-Wei Zhang[2]

[1]School of Mechanical and Aerospace Engineering, Nanyang Technological University, 639798 Singapore, Singapore

[2]Institute of High Performance Computing, Agency for Science, Technology and Research, 138632 Singapore, Singapore

[3]Institute for Metals Superplasticity Problems, Russian Academy of Sciences, 450001 Ufa, Russia

[4]National Research Tomsk State University, 634050 Tomsk, Russia

**Email:** *caiy@ihpc.a-star.edu.sg, † kzhou@ntu.edu.sg and ‡zhangyw@ihpc.a-star.edu.sg

**Abstract**

The poor structural stability of phosphorene in air was commonly ascribed to humidity and oxygen molecules. Recent exfoliation of phosphorene in deoxygenated water promotes the need to re-examine the role of $H_2O$ and $O_2$ molecules. Considering the presence of high population of vacancies in phosphorene, we investigate the interaction of $H_2O$ and $O_2$ molecules with vacancy-contained phosphorene using first-principles calculations. In contrast to the common notion that physisorbed molecules tend to have a stronger adsorption at vacancy sites, we show that $H_2O$ has nearly the same adsorption energy at the vacancy site as that at the perfect one. Charge transfer analysis shows that $O_2$ is a strong electron scavenger, which transfers the lone-pair electrons of the phosphorus atoms to the $2\pi^*$ antibonding orbital of $O_2$. As a result, the barrier for the O-O bond splitting to form O-P bonds is reduced from 0.81 eV at the perfect site to 0.59 eV at the defect site, leading to an about 5000 faster oxidizing rate at the defect site than at the perfect site at room temperature. Hence, our work reveals that the vacancy in phosphorene shows a stronger oxygen affinity than the perfect phosphorene lattice site. Structural degradation of phosphorene due to oxidization may occur rapidly at edges and grain boundaries where vacancies tend to agglomerate.

# Introduction

Phosphorene, the monolayer honeycomb structure of black phosphorus, has recently attracted great attention [1–5] due to its direct finite band gap [6, 7], intriguing chemical [8–11] and optical properties [12, 13]. Owing to strong quantum confinement, phosphorene exhibits a thickness-dependent work function and band gap (0.39 eV for bulk and 1.52 eV for monolayer) [14], ideal for infrared optoelectronics applications. In addition, its high carrier mobility [3] of 1000 cm$^2$/V·s at room temperature implies its promising applications in transistors and other nanoelectronic devices. Furthermore, its asymmetric electronic and phononic conduction [15, 16], negative Poisson's ratio [17] and highly flexible structure [18], also suggest that it is a promising material for thermoelectric and mechanoelectrical applications [19–24].

Clearly, those applications require our ability to massively produce large-area and high-quality phosphorene. Unlike other two-dimensional (2D) materials, such as graphene, boron nitride, and MoS$_2$, phosphorene still cannot be fabricated by chemical vapor deposition (CVD) owing to its relatively chemically reactive character. Currently, high-quality phosphorene sheets can only be obtained either by pressure-induced phase transformation [25] or by mechanical exfoliation [2], which, in general, suffers from poor scalability. In addition, liquid phase exfoliation, a conventional method for massive production of other 2D materials, has to be properly tailored in order to avoid the structural degradation of phosphorene arising from water and oxygen molecules [26–29]. To meet this requirement, anhydrous organic solvents have been used. Nevertheless, this method has some disadvantages, such as a limited exfoliation yield and suboptimal flake size distribution of phosphorene compared with other 2D materials exfoliated in aqueous solutions [1, 30, 31]. More recently, a scalable, high-yield and environmentally benign approach was demonstrated via chemical exfoliation of phosphorene [32]. Although water is generally considered as a direct cause of structural degradation of phosphorene, interestingly, this method is actually based on aqueous dispersion in deoxygenated water. Clearly, the generally assumed detrimental role of water needs to be reexamined, and the protective mechanism of deoxygenated water and the interactions of water and O$_2$ with phosphorene need to be understood.

Another important concern of phosphorene is related to its intrinsic instability arising from its phosphorus vacancies. It is known that for atomically thin 2D materials, lattice imperfections can be easily introduced during fabrication or intentionally produced via electron beam or other high-energy particle excitations [33–35]. For phosphorene, this issue seems to be even more critical since the atomic vacancies in phosphorene are calculated to be easily formed and abundant at ambient condition with their much lower formation energy (1.65 eV) than other 2D materials [36]. In addition, vacancies in phosphorene are highly mobile with an ultralow diffusion barrier of 0.30 eV compared of 1.39 eV for vacancy in graphene [36]. Such itinerant vacancies may greatly affect the stability of phosphorene in air with respect to the interaction with environmental molecules. While most of the previous studies only focus on the effects of environmental molecules absorbed on perfect phosphorene [37–42], the interactions of defected phosphorene with environmental molecules have not been considered so far.

In the present work, using first-principle calculations, we investigate the effect of absorption of $H_2O$ and $O_2$ molecules on the electronic structures of perfect, mono-vacancy (MV), and di-vacancy (DV)-contained phosphorene. It is found that unlike graphene and $MoS_2$, where defects enhance the adsorption of molecules at defective sites [38, 43], the vacancy-contained phosphorene shows almost the same chemical susceptibility to water as perfect phosphorene, due to their comparable energy release during adsorption, whereas for $O_2$, the presence of DV can greatly promote its adsorption. Since vacancies in phosphorene are abundant and itinerant [36], as a consequence, oxygen molecules may be easily trapped at those defect sites. We find that the vacancies have significant effects on the oxidation of phosphorene with a 5000 faster oxidizing rate at the defect site than at the perfect site. As the vacancies tend to accumulate at the edges and grain boundaries, structural failure is highly likely to initiate at these sites. We predict that passivating the vacancies should be an effective strategy to promote the stability of phosphorene. Moreover, vacancies in phosphorene are found to be able to modulate the charge transfer between water and $O_2$ molecules and phosphorene. The findings revealed here may render new ways to protect phosphorene from structure degradation and control the polarity and concentration of charge carriers in phosphorene.

## Computational methods

Density functional theory (DFT) calculations were performed by using VASP [44] packages. Perdew−Burke−Ernzerhof (PBE) [45] exchange-correlation functionals under the generalized gradient approximation (GGA) were selected. The Van der Waals (vdW) corrected functional with Becke88 optimization (optB88) [46] was used for treating the dispersive interactions during the noncovalent chemical functionalization of phosphorene with small molecules. All the structures were fully relaxed until the total energy and atomic forces were smaller than $10^{-5}$ eV and 0.01 eV/Å, respectively. The effects of MV and DV in phosphorene were considered by removing one or two phosphorus atoms in a $4 \times 5 \times 1$ supercell (80 phosphorus atoms). Periodic boundary conditions were applied in the two in-plane transverse directions, together with a vacuum space with a thickness of 20 Å. For all the considered cases, we chose the energy cutoff of 400 eV and the first Brillouin zone was sampled with a $10 \times 10 \times 1$ k-mesh grid. The absorption energy ($E_a$) of a molecule on perfect and vacancy-contained phosphorene surfaces was calculated as $E_a = E_{Mol+P} - E_P - E_{Mol}$, where $E_{Mol+P}$, $E_P$ and $E_{Mol}$ are the energies of the molecule adsorbed phosphorene, phosphorene surface, and molecule, respectively. The reaction barriers are calculated by using the climbing image nudged elastic band (CI-NEB) method.

## Results and discussions

**Electronic structure of vacancy-contained phosphorene**

According to the band structure shown in figure 1(a) (lower panel), perfect phosphorene is a direct semiconductor with a band gap of 0.88 eV, which is consistent with previous studies [18, 38, 39, 47, 48]. Note that this value is grossly underestimated due to the well-known deficiency in GGA. Similar to graphene, there exist many phases of MV and DV in phosphorene [36, 49–51]. Herein we only examine the lowest energy configuration of MV, which consists of pentagon-nonagon (59) rings as shown in figure 1(b) (upper panel), and that of DV, which consists of pentagon-heptagon-pentagon-heptagon (5757) rings as shown in figure 1(c) (upper panel).

For the 59 MV, removal of a phosphorus atom from perfect phosphorene creates unpassivated atoms and dangling bonds in the defect core. While the MV-contained phosphorene exhibits essentially the same band gap as perfect phosphorene, there is a significant readjustment of band lines according to figure 1(b) (lower panel). A partially occupied defect band, which crosses the Fermi level, appears at about 0.01 eV above the valence band maximum (VBM) of the host phosphorene, suggesting easy production of hole states (p-type conduction) even upon moderate thermal excitations. Although MV-contained phosphorene still possesses a direct band gap, the VBM of the host phosphorene shifts from $\Gamma$ point in the perfect case to Y point. This change in the band structure may affect the optical emission efficiency of phosphorene.

For the DV-contained phosphorene, it is found that the 5757 DV defect shifts the VBM and conduction band minimum (CBM) downward and upward, respectively, which leads to an increase in the band gap up to 1.04 eV (figure 1(c)). This increase may arise from the large lattice distortion and local strain induced by the DV [36]. Unlike MV defect, there is no defect state in the band gap for the DV-contained phosphorene due to the absence of dangling bond and the full passivation of atoms. In addition, similar to the MV case, the VBM shifts from $\Gamma$ to Y point and there is a direct-to-indirect band gap transition upon the introduction of 5757 DV. Such an increase in the band gap and direct-indirect band gap transition could be detectable in optical spectrum, and blue shifts of the emission and adsorption peaks may be used to corroborate the presence of the 5757 DV defect.

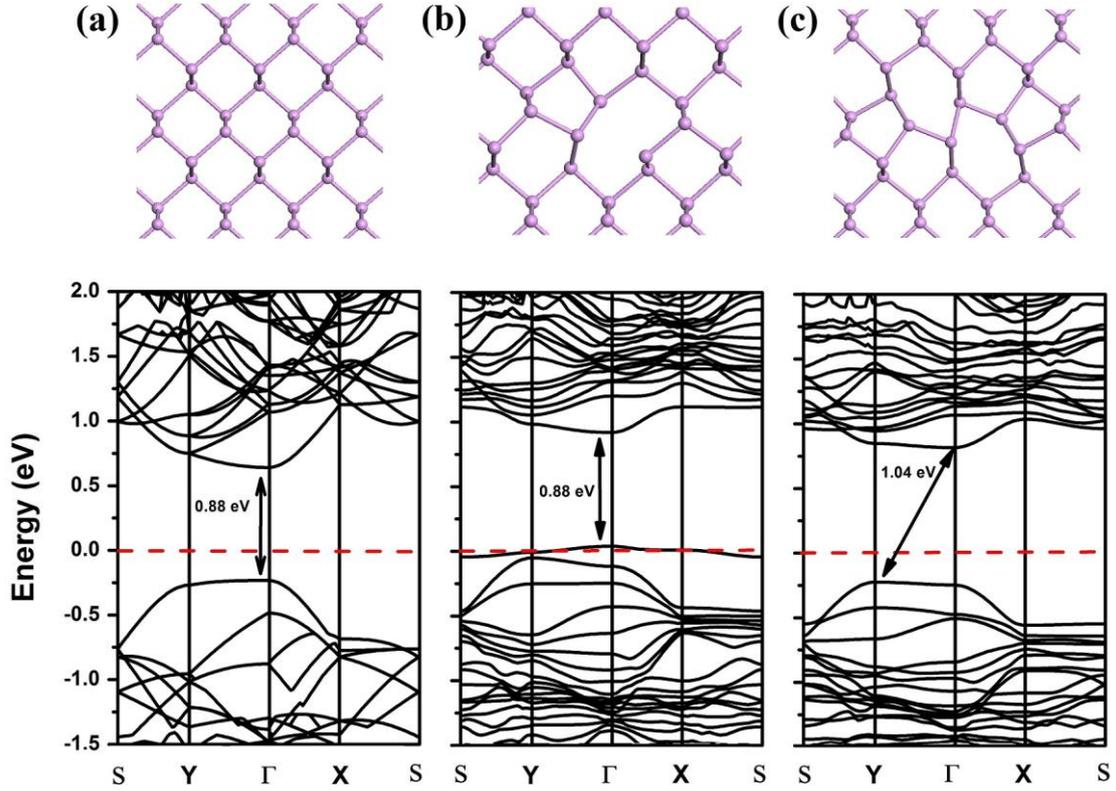

**Figure 1.** Atomic configurations (upper panel) and band structure (lower panel) of phosphorene (a) perfect, (b) with MV defect, (c) with 5757 DV defect. Red dashed lines indicate the Fermi level.

**Physisorption of $H_2O$ and $O_2$ molecules above vacancies**

We next consider the physisorption of the $H_2O$ and $O_2$ molecules above the phosphorus-deficient phosphorene. For each molecule, we have examined several possible absorption positions on perfect and defected phosphorene. All subsequent calculations on the electronic properties and energetics are based on the lowest-energy configuration and the energetics data of the $H_2O$ and $O_2$ adsorptions are compiled in Table 1.

The three lowest energy configurations for $H_2O$ and $O_2$ physical adsorptions on perfect, MV and DV-contained phosphorene are shown in figures 2 and 3, respectively. For the most stable binding configurations of $H_2O$ adsorbed on perfect phosphorene (figure 2(a)), one of the O−H bonds is oriented parallel to the surface along the armchair direction while the other is nearly normal to the surface. The in-plane O−H bond is located directly above the ridge of phosphorene. The distance from the molecule to the surface is 3.01 Å and the binding energy $E_a$ is −0.187 eV, which is consistent with previous work on phosphorene [37]. For the most stable binding configuration of $H_2O$ adsorption on the MV defect (figure 2(d)), both of the two O−H bonds are oriented nearly parallel to the surface and located directly above the MV position. The distance from the molecule to the surface is 2.42 Å and the binding energy $E_a$ is −0.193 eV. Figure 2(j) shows the lowest-energy geometry of $H_2O$ adsorbed on phosphorene with the DV defect, where the $H_2O$ is located above one of the pentagon rings of the 5757 defect with the adsorption height of 2.66 Å and the binding energy $E_a$ of −0.205 eV.

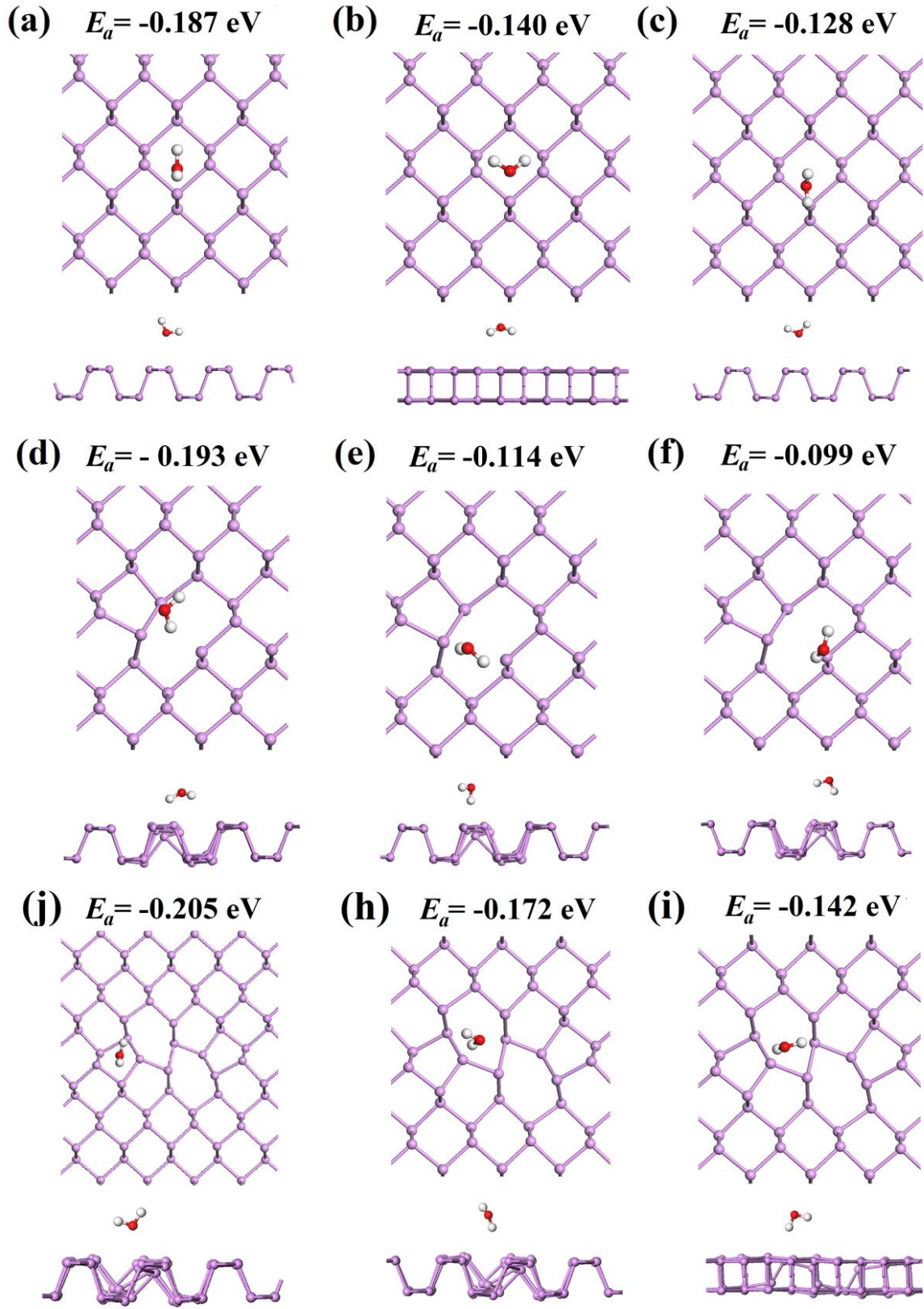

**Figure 2.** Top and side views of the examined possible absorption configurations of H$_2$O molecule adsorbed on phosphorene. (a), (b) and (c) for perfect, (d), (e) and (f) with MV defect, (j), (h) and (i) with DV defect. The balls in blue and red and white represent phosphorus, oxygen and hydrogen atoms, respectively. The lowest-energy configurations of H$_2$O molecule adsorbed on phosphorene are shown in (a), (d) and (j) for perfect, with MV defect and with DV defect, respectively.

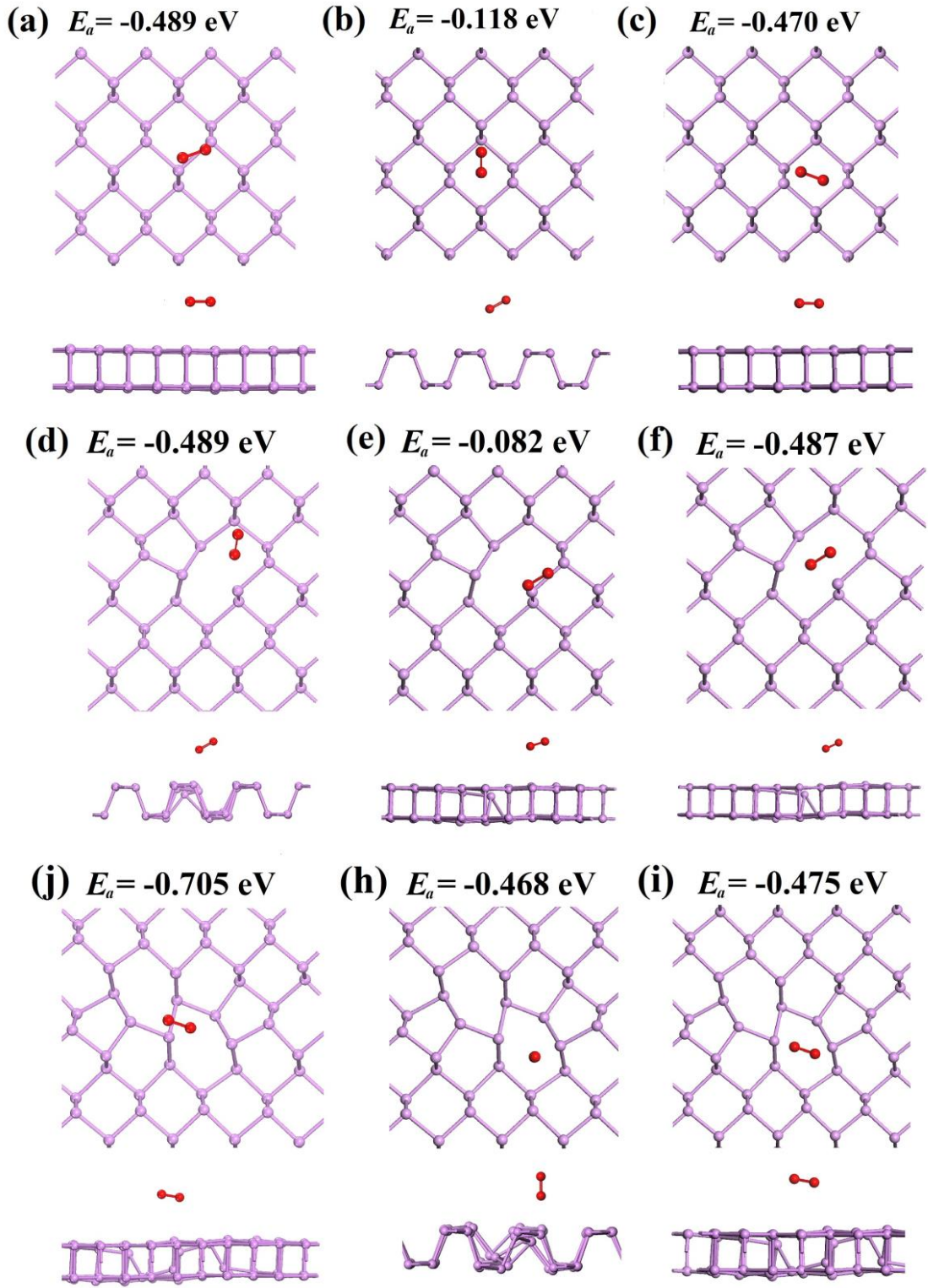

**Figure 3.** Top and side views of the examined possible absorption configurations of O₂ molecule adsorbed on phosphorene. (a), (b) and (c) perfect, (d), (e) and (f) with MV defect, (j), (h) and (i) with DV defect. The balls in blue and red represent phosphorus and oxygens atoms, respectively. The lowest-energy configurations of O₂ molecule adsorbed on phosphorene are shown in (a), (d) and (j) for perfect, with MV defect and with DV defect, respectively.

For the most stable binding configuration of $O_2$ adsorbed on perfect phosphorene (figure 3(a)), the O−O bond is oriented parallel to the surface along the armchair direction and located directly above the ridge. The distance from the molecule to the surface is 2.80 Å and the binding energy $E_a$ is −0.489 eV. The most stable configuration for the $O_2$ molecule absorbed on the MV defect is presented in figure 3(d), where the O−O bond is located directly above the MV position, tilting about 45° away from the surface. The distance from the molecule to the surface is 2.94 Å and the binding energy $E_a$ is −0.489 eV. For the most stable binding configurations of $O_2$ adsorbed on the DV defect (figure 3(j)), the O−O bond deviates slightly from the in-plane surface and is located directly above the central P−P bond shared by the two neighboring heptagons. The distance from the molecule to the surface is 3.02 Å and the binding energy $E_a$ is −0.705 eV. The O-O bond length of the isolated molecule changes from 1.22 Å to 1.25, 1.24, and 1.24 Å, upon adsorption on perfect, MV, and DV-contained phosphorene, respectively. This elongation of the O-O bond length signifies a strong electron transfer between the substrate and the $O_2$ molecule, and the transferred charges mostly occupy the $2\pi^*$ antibonding orbital. Therefore, the O-O bond is weakened even for a physisorbed $O_2$ molecule on phosphorene, and as a result, the energy for the O-O bond splitting is lowered, explaining the high affinity of phosphorene to oxygen.

Interestingly, in contrast to the common notion that defects in 2D materials generally have a higher chemical affinity to adsorbates, our results show that the presence of MV has almost negligible effect on the binding energy $E_a$ of $H_2O$ and $O_2$ compared with the adsorption on perfect surface. A possible underlying reason is that the defect states are well self-passivated due to the highly puckered structure of phosphorene since the atoms in the defect core cross two neighboring ridges and tend to have a stronger interaction and hybridization than other planar 2D materials like graphene and $MoS_2$. The above scenario is consistent with previous study showing that the defects in phosphorene are nearly electronically inert [52]. For the DV defect, it can only slightly enhance the physisorption of $H_2O$ molecule (with $E_a$ from -0.187 eV for perfect case to -0.205 eV for DV case) but greatly promote the adsorption of $O_2$ molecule (with $E_a$ from -0.489 eV for perfect case to -0.705 eV for DV case). The promoted interaction may be due to the large lattice distortion and bond deformation around the DV core. Our study suggests that the vacancy-contained phosphorene shows almost the same affinity to the water molecules from the thermodynamics point of view due to the comparable energy release with the physisorption above the perfect lattice.

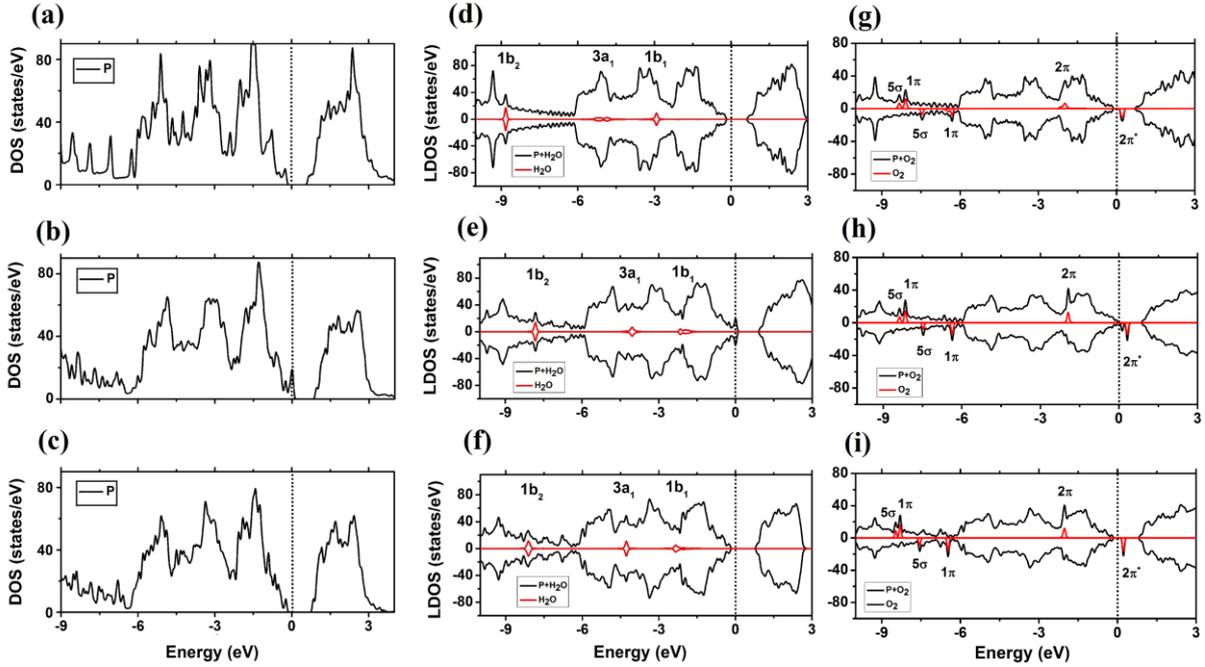

**Figure 4.** DOS structure of phosphorene: (a) perfect, (b) with MV defect, and (c) with DV defect. LDOS of $H_2O$ and $O_2$ on phosphorene: (d), (g) perfect, (e), (h) with MV defect, and (f), (i) with DV defect. Spin-up and -down bands for $H_2O$ and $O_2$ are symmetrical and shown by the red lines, while black lines represent the total DOS. Dashed lines indicate the Fermi level.

**Electronic structure and states alignment**

Figure 4(a)−(c) presents the density of states (DOS) of perfect phosphorene, and phosphorene with MV and DV defects, respectively. It is shown that a MV defect can cause an enhancement in the electronic states around the top of the valence band as reflected by the increase in the peak intensity in the local density of states (LDOS) in figure 4(b) compared with that of perfect phosphorene in figure 4(a). This is attributed to the newly formed defect states above the VBM as shown in the band structure of figure 1. For the 5757 DV defect, as shown in figure 4(c), the DOS profile is quite similar to that of perfect phosphorene, and there are no defect states within the band gap.

In contrast, for the $H_2O$ physisorption, no additional electronic state within the fundamental band gap is formed for either perfect or defected phosphorene (figure 4(d)−(f)). The value of the respective band gap for perfect, MV and DV-contained phosphorene is almost the same as pristine phosphorene. However, the presence of vacancies on the surface significantly affects the alignment of the molecular levels of $H_2O$ with respect to those of phosphorene. The three highest occupied molecular orbitals (HOMO) of the H$_2$O molecule, named according to the irreducible representation of the point group of H$_2$O, are 1b$_1$ (HOMO), 3a$_1$ (HOMO-1), and 1b$_2$ (HOMO-2). All these levels are greatly upwardly shifted by around 1 eV in the MV- and DV-contained phosphorene. This readjustment of alignment of the molecular levels is a clear indication of a different amount of charge transfer and different interactions between water and phosphorene. Interestingly, for H$_2$O adsorbed on perfect sheet, the 3a$_1$ orbital is the most broadened one due to its favored orbital mixing with the P atom. The situation becomes different for the adsorption of MV and DV defects, where the 1b$_1$ state of the H$_2$O molecule is the most broadened one. This difference reflects the fact that H$_2$O is prone to have a different binding mechanism at the vacancy site compared with prefect one.

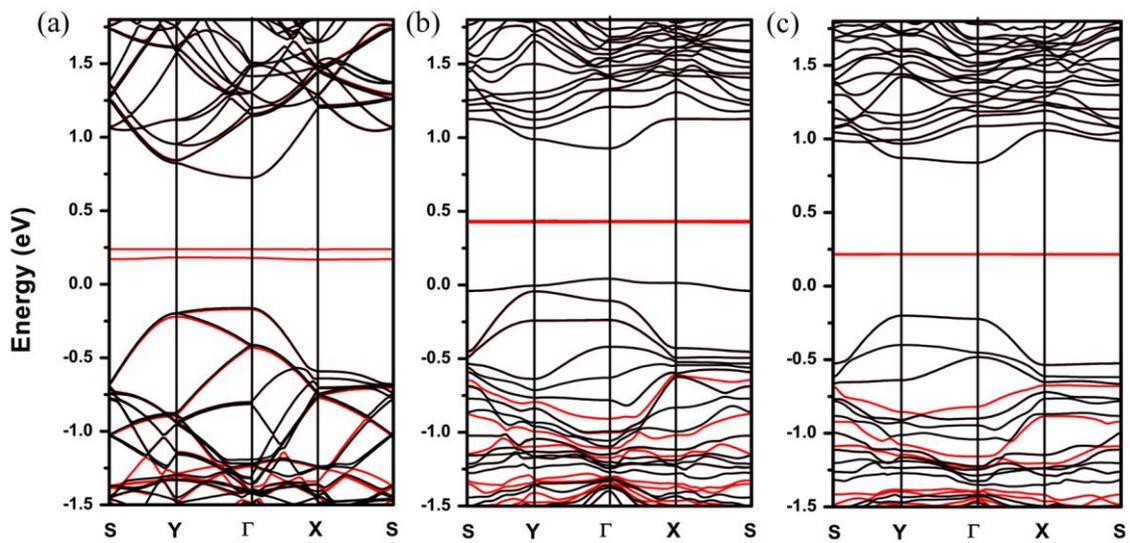

**Figure 5.** Band structure of O$_2$ on phosphorene: (a) perfect, (b) with MV defect, and (c) with DV defect. Energetic levels associated with the O$_2$ are plotted in red.

In contrast, for O$_2$ molecule, its physisorption can substantially modify the electronic structure of both perfect and defected phosphorene. Figure 4(g)−(i) shows the LDOS for perfect, MV and DV-contained phosphorene, respectively. The adsorption of O$_2$ induces additional states with HOMO being located in the proximity of the VBM region. For all the cases, the antibonding LUMO state (2π*, down) is located in the band gap of phosphorene above the Fermi level, while the HOMO state (2π, up) is slightly broadened for perfect and narrowed for MV and DV-contained phosphorene. Figure 5 shows the band structure of O$_2$-adsorbed phosphorene for the three cases. The spin triplet states (LUMO, 2π*) of O$_2$ remains unoccupied for all the cases with the degeneracy being strongly lifted for the perfect case. Unlike the case of $H_2O$ molecule absorption, the alignment of the energetic level of orbitals of $O_2$ with that of phosphorene is almost insensitive to the presence of vacancies. Therefore, the $O_2$ passivation of vacancies is able to induce trap states in the band gap of phosphorene, which is different from the case of sulfur vacancy in $MoS_2$, where the $O_2$ adsorption at the vacancy site can change the electronic nature of the vacancies from carrier-traps to electronically benign sites [38].

**Table 1.** Absorption energy ($E_a$), charge transfer ($\Delta q$) from $H_2O$ and $O_2$ molecules to phosphorene, and X−P bond length ($B_{X-P}$), where X represents the $H_2O$ or $O_2$ molecules. Note that a positive $\Delta q$ indicates the transfer of electrons from the molecules to phosphorene.

| Molecule | Phosphorene | $E_a$ (eV) | $\Delta q$ ($e$) | $B_{x-p}$ (Å) | Molecule |
|---|---|---|---|---|---|
| $H_2O$ | Perfect | -0.187 | 0.010 | 3.01 | donor |
|  | MV | -0.193 | 0.120 | 2.42 | donor |
|  | DV | -0.205 | 0.050 | 2.66 | donor |
| $O_2$ | Perfect | -0.489 | -0.036 | 2.80 | acceptor |
|  | MV | -0.489 | -0.030 | 2.94 | acceptor |
|  | DV | -0.705 | 0.010 | 3.02 | donor |

**Modulation of carrier density and charge transfer**

To analyze the electronic interaction between the $H_2O$ and $O_2$ molecules with phosphorene, we calculated the differential charge density (DCD) $\Delta\rho(\mathbf{r})$ defined as the difference between the total charge density of molecularly adsorbed phosphorene system subtracted by the sum of the charge densities of the isolated molecule and the naked phosphorene. To obtain the exact amount of transferred charge from the $H_2O$ or $O_2$ molecule, the plane-averaged DCD $\Delta\rho(z)$ along the normal direction (z) of the phosphorene sheet is calculated by integrating $\Delta\rho(\mathbf{r})$ within the basal plane at the z point. The amount of transferred charge at z point is given by $\Delta Q(z) = \int_{-\infty}^{z} \Delta\rho(z')dz'$. Based on the $\Delta Q(z)$ curves, the total amount of charge donated by the molecule is read at the interface between the molecule and the phosphorene, where $\Delta\rho(z)$ shows a zero value. The isosurface of $\Delta\rho(\mathbf{r})$ for the $H_2O$ molecule adsorbed on perfect phosphorene and phosphorene with MV and DV defects is depicted in figure 6(a)−(c), respectively. It is seen that there is a depletion of electrons in $H_2O$ molecule and an accumulation of electrons in the nearest P atoms of perfect surface (figure 6(a)), and the $H_2O$ molecule donates electrons to phosphorene with around 0.01 *e* per molecule. In the case MV defect, the donor ability of $H_2O$ molecule is increased and the total amount of transferred charge increases significantly up to 0.12 *e*. In case of DV defect, the total amount of transferred charge from $H_2O$ is 0.05 *e*. Due to the charge transfer from water to phosphorene, an effective dipole pointing toward vacuum should be established across the molecule-phosphorene interface. It is expected that the surface coverage of $H_2O$ molecules under humidity condition could decrease the work function of phosphorene layer due to the presence of the dipole layer, which in turn could affect the charge injection from the electrode to the channel layer and thus the device performance.

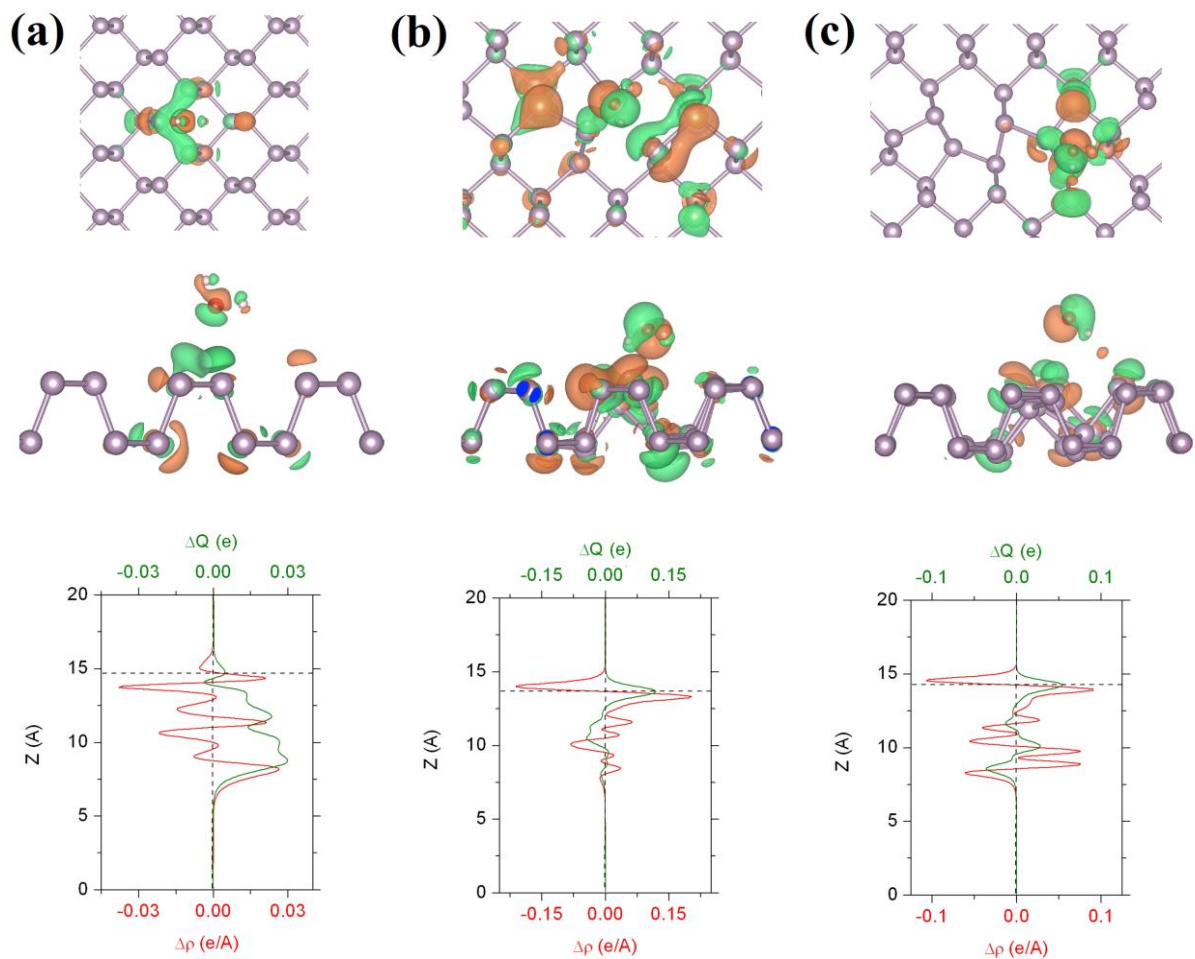

**Figure 6.** Charge redistribution for the $H_2O$ molecule absorbed on perfect (a), MV contained (b), and DV contained (c) phosphorene. Top and middle panels: Top and side views of the 0.02 Å$^{-3}$ DCD isosurface. The green (orange) color denotes depletion (accumulation) of electrons. Bottom panel: Plane-averaged differential charge density $\Delta\rho(z)$ (red line) and the amount of transferred charge $\Delta Q(z)$ (green line) between the $H_2O$ molecule and phosphorene.

Figure 7(a)−(c) presents the isosurface of Δρ(r) for the $O_2$ molecule adsorbed on perfect phosphorene, and phosphorene with MV and DV defects, respectively. It is found that $O_2$ accepts electrons from perfect phosphorene with around 0.035 $e$ per molecule. The MV defect slightly decreases the donor ability of $O_2$ molecule with the total amount of charge transfer amounting to 0.03 $e$. In contrast, the DV defect receives a tiny charge transfer of 0.01 $e$ from the molecule, partly due to the fully compensated structure and weak dipole interaction. Therefore, the carrier density of phosphorene can be modulated by water molecules, oxygen molecules and vacancies.

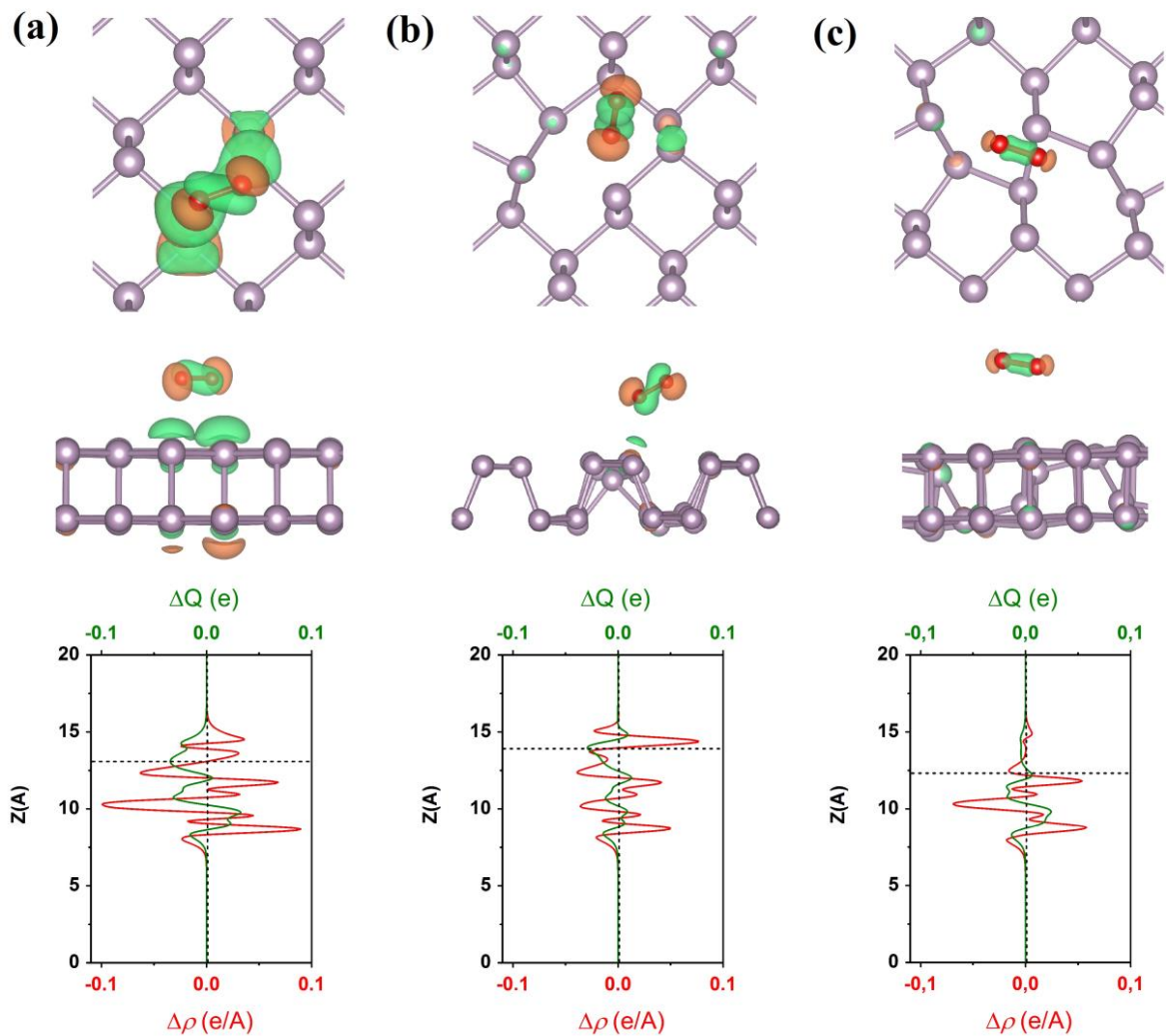

**Figure 7.** Charge redistribution for the $O_2$ molecule absorbed on perfect (a), MV contained (b), and DV contained (c) phosphorene. Top and middle panels: Top and side views of the 0.02 Å$^{-3}$ DCD isosurface. The green (orange) color denotes depletion (accumulation) of electrons. Bottom panel: Plane-averaged differential charge density Δρ(z) (red line) and the amount of transferred charge ΔQ(z) (green line) between the $O_2$ molecule and phosphorene.

**Effect of MV on the dissociation of O₂ molecule**

Experiments have shown that phosphorene can be easily oxidized in air condition largely due to the oxygen molecules [26–29]. However, the underlying mechanism of the kinetic process from gas $O_2$ molecule to form chemically bonded O-P species is still unclear. Recent work [37] on GaS and $MoS_2$ semiconductors has shown that most molecules, including $H_2O$, are only physisorbed on defects, while the $O_2$ molecule may reach chemisorbed state from the physisorbed state if the energy barrier is overcome. The present study shows that $H_2O$ molecule can only be physisorbed while $O_2$ molecule experiences an energy barrier from the physisorption to chemisorption on phosphorene. We find that this barrier can be strongly affected by the presence of vacancies in phosphorene. The detailed pathway from the initial state (IS), to the transition state (TS) and to the final state (FS) for oxidation of phosphorene by $O_2$ gas molecule on perfect and MV sites are shown in figure 8(a)−(c). The calculated energy barrier $E_b$ for the perfect case is 0.81 eV. From figure 8(b), it is seen that the presence of MV can significantly reduce the barrier to 0.59 eV. According to these results, a large amount of $O_2$ molecules in air is able to be physisorbed at room temperature. Our obtained results on the chemisorbed energies (4 eV per $O_2$) are in good agreement with a recent work [53].

According to the rate theory, the transition time from the physisorbed state to the chemisorbed state is $t \approx 1/\left(f \cdot e^{-E_b/k_b \cdot T}\right)$, where $E_b$ is the barrier, $k_b$ is the Boltzmann constant, $T$ is a temperature and $f$ is the attempt frequency, defined as $f = n \cdot v \cdot s_d$, where $n$ is $O_2$ density in air, $v$ is the speed, and $s_d$ can be taken as the square of lattice parameter. Hence, at the room temperature of 300 K, one atmospheric pressure, and $f$ of arround $10^8$ molecules/s, the time of $O_2$ molecule chemisorbtion on perfect phosphorene is $t \approx$ 109 hours. This value reduces to 1.33 min on the MV site, which is about 5000 times shorter. Thus, our work suggests that the oxidation rate is much higher at the vacancies than at the perfect sites and that phosphorene sheets with high-concentration vacancies can be more easily oxidized than vacancy-free phosphorene. Passivation and repairment of these vacancies in phosphorene should be effective in enhancing the chemical stability of phosphorene. However, the oxidation is also limited by the possible absorbed sites. The formation energy of P vacancy is 1.65 eV, and the concentration of the intrinsic vacancy estimated by $N_{host}*\exp(-1.65/kT)$, where $N_{host}$ is the total number of P atoms of the corresponding perfect lattice, is several orders of magnitude smaller than that of the host P sites. Hence, the oxidation rate of phosphorene is still largely dominated by the reaction at the perfect sites. Effects of vacancies tend to be more significant for small size phosphorene flakes which contain a large amount of edges with accumulated vacancies.

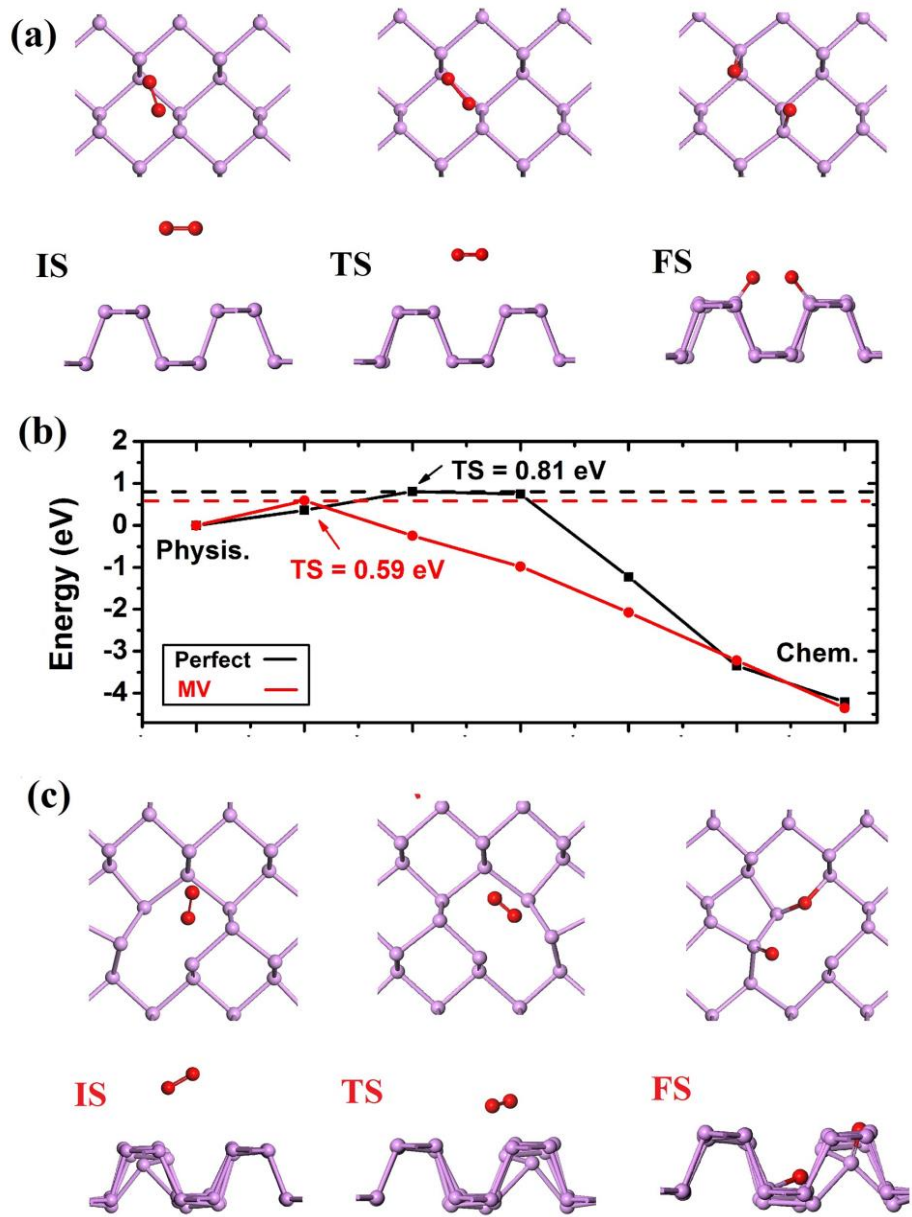

**Figure 8.** Atomic configurations from the physisorbed to the chemisorbed state in the dissociation process of $O_2$ molecule on (a) perfect (black line) and (c) MV (red line) phosphorene. P and O atoms are colored in purple and red, respectively. (b) Energetic profiles of the reaction pathway obtained from NEB calculations.

## Conclusions

By using first-principles calculations, we investigate the interaction of vacancy-contained phosphorene with $H_2O$ and $O_2$ molecules. It is found that different from other 2D materials, vacancy-contained phosphorene is almost inert to $H_2O$ with the adsorption energy being almost the same as that in perfect phosphorene. For both perfect and vacancy-contained phosphorene, $H_2O$ molecule does not introduce any defect states in the band gap while the frontier orbitals of $O_2$ molecule are aligned in the band gap of the VBM of the phosphorene. $O_2$ molecule increases hole carriers and serves as a good electron scavenger for adsorption above perfect phosphorene. Vacancy-modulated charge transfer from $H_2O$ and $O_2$ molecules may allow the modulation of the concentration and polarity of carriers in phosphorene. Finally, we investigate the kinetics of $O_2$ dissociation and find that the oxidation rate is around 5000 times faster in the vacancy site than the perfect site. Phosphorene samples with a large amount of vacancies should be more easily oxidized than those of low-vacancy contained phosphorene. The new understandings revealed here for the interactions of $O_2$ and $H_2O$ molecules with phosphorene may inspire new strategies to exfoliate and protect phosphorene.

## Acknowledgments

The authors gratefully acknowledge the financial support from the Ministry of Education, Singapore (Academic Research Fund TIER 1 - RG128/14), the Agency for Science, Technology and Research (A*STAR), Singapore, and the use of computing resources at the A*STAR Computational Resource Centre, Singapore. Sergey V. Dmitriev acknowledges financial support from the Russian Science Foundation grant N 14-13-00982.

## References

[1]. Yasaei P, Kuma B, Foroozan T, Wang C, Asadi M, Tuschel D, Indacochea J E, Klie R F and Khojin A S High-quality black phosphorus atomic layers by liquid-phase exfoliation 2015 *Adv. Mater.* **27** 1887−1892.

[2]. Castellanos-Gomez A et al. Isolation and characterization of few-layer black phosphorus. 2014 *2D Mater.* **1** 025001.

[3]. Liu H, Neal A, Zhu Z, Luo Z, Xu X, Tománek D and Ye P Phosphorene: An unexplored 2D semiconductor with a high hole mobility. 2014 *ACS Nano*. **8** 4033–4041.

[4]. Guan J, Zhu Z and Tománek D Tiling phosphorene. 2014 *ACS Nano*. **8** 12763–12768.

[5]. Jing Y, Zhang X and Zhou Z. Phosphorene: what can we know from computations? 2016 WIREs Comput. Mol. Sci. **6** 5–19.

[6]. Rodin A S, Carvalho A and Castro Neto A H Strain-induced gap modification in black phosphorus. 2014 *Phys. Rev. Lett.* **112** 176801.

[7]. Tran V, Soklaski R, Liang Y F and Yang L Layer-controlled band gap and anisotropic excitons in few-layer black phosphorus. 2014 *Phys Rev B*. **89** 235319.

[8]. Favron A, Gaufrès E, Fossard F, Phaneuf-L'Heureux A-L, Tang N, Lévesque P L, Loiseau A, Leonelli R, Francoeur S and Martel R Photooxidation and quantum confinement effects in exfoliated black phosphorus. 2015 *Nat Mater*. **14** 826–832.

[9]. Wang H, Yang X, Shao W, Chen S, Xie J, Zhang X., Wang J and Xie Y Ultrathin black phosphorus nanosheets for efficient singlet oxygen generation. 2015 *J. Am. Chem. Soc*. **137** 11376–11382.

[10]. Rahman M Z, Kwong C W, Davey K and Qiao S Z 2D phosphorene as a water splitting photocatalyst: fundamentals to applications. 2016 *Energy Environ. Sci.* **9** 709-728.

[11]. Gan Z X, Sun L L, Wu X L, Meng M., Shen J C and Chu P K Tunable photoluminescence from sheet-like black phosphorus crystal by electrochemical oxidation. 2015 *Appl. Phys. Lett*. **107** 021901.

[12]. Qiao J, Kong X, Hu Z, Yang F and Ji W High-mobility transport anisotropy and linear dichroism in few-layer black phosphorus. 2014 *Nat. Commun.* **5** 4475.

[13]. Jing Y, Tang Q, He P, Zhou Z and Shen P. Small molecules make big differences: molecular doping effects on electronic and optical properties of phosphorene. 2015 *Nanotechnology* **26** 095201.

[14]. Cai Y, Zhang G and Zhang Y-W Layer-dependent band alignment and work function of few-layer phosphorene. 2014 *Scientific Reports*. **4** 6677.

[15]. Cai Y, Ke Q, Zhang G, Feng Y P, Shenoy V B and Zhang Y-W Giant phononic anisotropy and unusual anharmonicity of phosphorene: Interlayer coupling and strain engineering. 2015 *Adv. Funct. Mater.* **25** 2230−2236.

[16]. Fei R and Yang L Strain-engineering the anisotropic electrical conductance of few-layer black phosphorus. 2014 *Nano Lett.* **14** 2884−2889.

[17]. Jiang J-W and Park H S Negative Poisson's ratio in single-layer black phosphorus. 2014 *Nat. Commun.* **5** 4727.

[18]. Kistanov A A, Cai Y, Zhou K, Dmitriev S V and Zhang Y-W Large electronic anisotropy and enhanced chemical activity of highly rippled phosphorene. 2016 *J. Phys. Chem. C.* **120** 6876−6884.

[19]. Li L, Yu Y, Ye G J, Ge Q, Ou X, Wu H, Feng D, Chen X H and Zhang Y Black phosphorus field-effect transistors. 2014 *Nat. Nanotechnol.* **9** 372–377.

[20]. Zhao S, Kang W and Xueac J The potential application of phosphorene as an anode material in Li-ion batteries. 2014 *J. Mater. Chem. A.* **2** 19046−19052.

[21]. Xia F, Wang H and Jia Y Rediscovering black phosphorus as an anisotropic layered material for optoelectronics and electronics. 2014 *Nat. Commun.* **5** 4458.

[22]. Du Y, Liu H, Deng Y and Ye P D Device perspective for black phosphorus field-effect transistors: Contact resistance, ambipolar behavior, and scaling. 2014 *ACS Nano.* **8** 10035−10042.

[23]. Quereda J, San-Jose P, Parente V, Vaquero-Garzon L, Molina-Mendoza A J, Agraït N, Rubio-Bollinger G, Guinea F, Roldán R and Castellanos-Gomez A Strong modulation of optical properties in black phosphorus through strain-engineered rippling. 2016 *Nano Lett.* **16** 2931–2937.

[24]. Buscema M, Groenendijk D J, Steele G A, van der Zant H S J and Castellanos-Gomez A Photovoltaic effect in few-layer black phosphorus PN junctions defined by local electrostatic gating. 2014 *Nat. Commun.* **5** 4651.

[25]. Li X, Deng B, Wang X, Chen S, Vaisman M, Karato S, Pan G, Lee M L, Cha J and Wang H Synthesis of thin-film black phosphorus on a flexible substrate. 2015 *2D Materials.* **2** 031002.

[26]. Kang J, Wood J D, Wells S A, Lee J-H, Liu X, Chen K-S and Hersam M C Solvent exfoliation of electronic-grade, two-dimensional black phosphorus. 2015 *ACS Nano.* **9** 3596–3604.

[27]. Serrano-Ruiz M, Caporali M, Ienco A, Piazza V, Heun S and Peruzzini M The role of water in the preparation and stabilization of high-quality phosphorene flakes. 2016 *Advanced Materials Interfaces*. **3** 1500441.

[28]. Xu J-Y, Gao L-F, Hu C-X, Zhu Z-Y, Zhao M, Wang Q and Zhang H-L Preparation of large size, few-layer black phosphorus nanosheets via phytic acid-assisted liquid exfoliation. 2016 *Chem. Commun*. **52** 8107-8110.

[29]. Passaglia E, Cicogna F, Lorenzetti G, Legnaioli S, Caporali M, Serrano-Ruiz M, Ienco A and Peruzzini M Novel polystyrene-based nanocomposites by phosphorene dispersion. 2016 *RSC Adv*. **6** 53777-53783.

[30]. Hanlon D et al. Liquid exfoliation of solvent-stabilized few-layer black phosphorus for applications beyond electronics. 2015 *Nat Commun.* **6** 8563.

[31]. Zhao W, Xue Z, Wang J, Jiang J, Zhao X and Mu T Large-scale, highly efficient, and green liquid-exfoliation of black phosphorus in ionic liquids. 2015 *ACS Applied Materials & Interfaces*. **7** 27608-27612.

[32]. Kang J, Wells S A, Wood J D, Lee J-H, Liu X, Ryder C R, Zhu J, Guest J R, Husko C A and Hersam M C Stable aqueous dispersions of optically and electronically active phosphorene. 2016 *PNAS*. doi:10.1073/pnas.1602215113.

[33]. Song Y L, Zhang Y, Zhang J M, Lu D B and Xu K W First-principles study of the structural and electronic properties of armchair silicene nanoribbons with vacancies. 2011 *Journal of Molecular Structure*. **990** 75–78.

[34]. Yazyev O V and Helm L Defect-induced magnetism in graphene. 2007 *Phys. Rev. B*. **75** 125408.

[35]. Ataca C, Sahin H, Akturk E and Ciraci S Mechanical and electronic properties of $MoS_2$ nanoribbons and their defects. 2011 *J. Phys. Chem. C*. **115** 3934–394.

[36]. Cai Y, Ke Q, Zhang G, Yakobson B I and Zhang Y-W https://arxiv.org/abs/1607.04523.

[37]. Cai Y, Ke Q, Zhang G and Zhang Y-W Energetics, charge transfer, and magnetism of small molecules physisorbed on phosphorene. 2015 *J. Phys. Chem. C*. **119** 3102−3110.

[38]. Liu Y, Stradins P and Wei S H Air passivation of chalcogen vacancies in two-dimensional semiconductors. 2016 *Angew. Chem. Int. Ed.* **55** 965–968.

[39]. Hu T and Dong J Geometric and electronic structures of mono- and di-vacancies in phosphorene. 2015 *Nanotechnology*. **26** 065705.

[40]. Wang G, Slough W J, Pandey R and Karna S P Degradation of phosphorene in air: understanding at atomic level. 2016 *2D Mater*. **3** 025011.

[41]. Utt K L, Rivero P, Mehboudi M, Harriss E, Borunda M F, SanJuan A A P and Barraza-Lopez S Intrinsic defects, fluctuations of the local shape, and the photo-oxidation of black phosphorus. 2015 *ACS Cent. Sci.* **1** 320–327.

[42]. Wood J D, Wells S A, Jariwala D, Chen K-S, Cho E, Sangwan V K, Liu X, Lauhon L J, Marks T J and Hersam M C Effective passivation of exfoliated black phosphorus transistors against ambient degradation. 2014 *Nano Lett.* **14** 6964–6970.

[43]. Banhart F, Kotakoski J and Krasheninnikov A V Structural defects in graphene. 2011 *ACS NANO*. **5** 26–41.

[44]. Kresse G and Furthmüller J Efficient iterative schemes for ab initio total-energy calculations using a plane-wave basis set. 1996 *Phys. Rev. B: Condens. Matter Mater. Phys.* **54** 11169.

[45]. Perdew J P, Burke K and Ernzerhof M Efficient iterative schemes for ab initio total-energy calculations using a plane-wave basis set. 1996 *Phys. Rev. Lett.* **77** 3865−3868.

[46]. Becke A D Density-functional exchange-energy approximation with correct asymptotic behavior. 1988 *Phys. Rev. A: At., Mol., Opt. Phys.* **38** 3098.

[47]. Hu W and Yang J Defects in phosphorene. 2015 *J. Phys. Chem. C*. **119** 20474−20480.

[48]. Srivastava P, Hembram K P S S, Mizuseki H, Lee K-R, Han S S and Kim S Tuning the electronic and magnetic properties of phosphorene by vacancies and adatoms. 2015 *J. Phys. Chem. C.* **119** 6530−6538.

[49]. Li X B, Guo P, Cao T F, Liu H, Lau W M and Liu L M Structures, stabilities, and electronic properties of defects in monolayer black phosphorus. 2015 *Scientific Reports*. **5** 10848.

[50]. Wang V, Kawazoe Y and Geng W T Native point defects in few-layer phosphorene. 2015 *Phys. Rev. B*. **91** 045433.

[51]. Guo Y and Robertson J Vacancy and doping states in monolayer and bulk black phosphorus. 2015 *Scientific Reports*. **5** 14165.

[52]. Liu Y, Xu F, Zhang Z, Penev E S and Yakobson B I Two-Dimensional mono-elemental semiconductor with electronically inactive defects: the case of phosphorus. 2014 *Nano Lett*. **14** 6782–6786.

[53]. Ziletti A, Carvalho A, Campbell D K, Coker D F and Castro Neto A H Oxygen defects in phosphorene. 2015 *Phys. Rev. Lett.* **114** 046801.